\documentclass[aps,prl,amsmath,amssymb,twocolumn,superscriptaddress,showpacs]{revtex4-1}
\usepackage{graphicx}
\usepackage{hyperref}
\hypersetup{
	colorlinks=true,       % false: boxed links; true: colored links
	linkcolor=blue,          % color of internal links
	citecolor=blue,        % color of links to bibliography
	filecolor=magenta,      % color of file links
	urlcolor=blue         
}
\usepackage{natbib}

\begin{document}
%\doi{10.1080/0950034YYxxxxxxxx}
%\issn{1362-3044}
%\issnp{0950-0340}  \jvol{00} \jnum{00} \jyear{2012} \jmonth{10 March}

\title{All-atomic generation and noise-quadrature filtering of squeezed vacuum in hot Rb vapor}

\author{Travis Horrom}
\affiliation{Department of Physics, The College of William and Mary, Williamsburg, VA 23187, USA}

\author{Gleb Romanov}
\affiliation{Department of Physics, The College of William and Mary, Williamsburg, VA 23187, USA}

\author{Irina Novikova}
\affiliation{Department of Physics, The College of William and Mary, Williamsburg, VA 23187, USA}

\author{Eugeniy E. Mikhailov}
\affiliation{Department of Physics, The College of William and Mary, Williamsburg, VA 23187, USA}

\begin{abstract}

With our all-atomic squeezing and filtering setup,
we demonstrate control over the noise amplitudes and manipulation of the
frequency-dependent squeezing angle of a squeezed vacuum quantum state
by  passing it  through  an atomic medium with electromagnetically induced transparency
(EIT).  We  generate low  sideband
frequency squeezed vacuum using the  polarization self-rotation effect in a
hot Rb vapor cell,  and input it into a second atomic  vapor subject to EIT
conditions. We use the frequency-dependent  absorption of the EIT window to
demonstrate an example  of squeeze amplitude attenuation  and squeeze angle
rotation of  the quantum noise quadratures  of the squeezed probe.  These studies
have implications for  quantum memory and storage as  well as gravitational
wave interferometric detectors.

\end{abstract}

\maketitle
\section{Introduction}

Squeezed light is a quantum-mechanical  state of electromagnetic field where the nonclassical photon
statistics allow  quadrature uncertainties  to be reduced  below the
shot
noise level, also referred  as the  standard quantum limit (SQL). 
Since the initial observation by Slusher~{\it et. al}~\cite{slusher85prl},
squeezed   light   and
squeezed vacuum  states  have  been  successfully  implemented  to  improve
precise  measurements in spectroscopy~\cite{polzik92prl}, 
interferometry~\cite{caves1981prd, Chickarmane96pra, kimble2002prd,
goda08nph_quantum-enhanced_gw_detector}, and
magnetometry~\cite{mitchel2010prl_sqz, mikhailov2012sq_magnetometer}.
Squeezing
has been intensely studied in  relation to quantum information and quantum
measurement  protocols~\cite{polzikRMP10,  euromemory, novikova2012review,
walls1983nature}.
Studies of quantum memory realizations based on the electromagnetically
induced transparency
(EIT) effect~\cite{harris'97pt} lead to  several experiments exploring
the propagation and storage of a  squeezed state
with EIT~\cite{akamatsu2004prl, akamatsu_ultraslow_2007, lvovsky08prl_sq_eit,
lvovsky09njp_squeeziong_eit, hondaPRL2008_sq_storage_in_Rb, kozumaOE07, kozumaPRA10} in
connection to quantum memory applications.

However,  without tools to effectively  manipulate the noise
properties of squeezed states, their applications are somewhat limited.
In particular, enhancing the sensitivity  of gravitational wave detectors
based on
interferometers (such as LIGO) with squeezing requires frequency-dependent
squeezing angles,
or at least degrees of squeezing which can combat the
greater  effects of radiation
pressure noise at lower detection frequencies~\cite{kimble2002prd}.
Such a manipulation of squeezing is possible with an optical
cavity~\cite{schnabel2005pra_squeezing_control}, but it requires
a narrow cavity linewidth below 1~kHz, set by LIGO internal cavities.
Such a high finesse cavity  would be large and bulky (on the order of 10  meters or larger) even with ultra-high
reflecting  mirrors.

It   was   suggested  by   Mikhailov   et   al.  ~\cite{mikhailov2005pra_eit_rot} that narrow optical
transmission resonances arising from coherent interaction with atoms 
(known   as
electromagnetically  induced transparency  (EIT) resonances)  could be  used to  create
frequency-dependent  filters   for  the  quadrature  noise   amplitude  and
angle  in  squeezed   light  states  to  be  used   in  gravitational  wave
detection.   Examples  of the amplitude
filtering effect  of squeezed  vacuum with EIT have  been 
observed~\cite{akamatsu2004prl,   akamatsu_ultraslow_2007,  lvovsky08prl_sq_eit,
lvovsky09njp_squeeziong_eit, hondaPRL2008_sq_storage_in_Rb, kozumaOE07, kozumaPRA10}. 
However, to the best of our knowledge the frequency-dependent squeezing angle
manipulation has not yet been demonstrated.

In the above experiments demonstrating quantum  noise filtering and atomic
memory,
the  squeezed vacuum  probes  were generated  using frequency doubling
nonlinear crystals  used in
optical  parametric  oscillators (OPOs).  This  method  has been  shown  to
generate  more than  11 dB  noise reduction~\cite{schnabel2010with11db}  at
1064~nm, however such  high squeezing has not been shown  to extend down to
the lower wavelengths required for EIT experiments due  to increased losses
in  nonlinear crystals. As a result, groups using  nonlinear crystal based
squeezers  for EIT experiments
with Rb atoms at 795~nm typically use around 3~dB noise suppressed squeezed
states~\cite{akamatsu2004prl, akamatsu_ultraslow_2007, lvovsky08prl_sq_eit,
lvovsky09njp_squeeziong_eit,   hondaPRL2008_sq_storage_in_Rb,  kozumaOE07}.
Crystal squeezing also  suffers from a high experimental complexity and
high laser power requirements.

In  this  paper,  we  demonstrate  the  EIT  noise amplitude  filtering 
and {\bf frequency-dependent} squeezing angle manipulation in
an   experimental   setup  using   only   atomic   vapors, for both
squeezing generation and manipulation.
Our squeezer utilizes the  polarization self-rotation  effect  in hot  Rb
vapor~\cite{MatskoNWBKR02, mikhailov2008ol}.  This squeezing  method offers
upwards of 3~dB  noise suppression~\cite{lezama2011pra} 
at low sideband frequencies  (200 Hz - 2 MHz)~\cite{mikhailov2012sq_magnetometer}, 
with  an economical all-atomic
experimental  design, using  low laser 
powers and neither nonlinear crystals nor optical cavities. 
We  also  observe  that  excess noise  may  couple  into  the
system  due to  the back  action of  light noise  onto the  atoms.
Understanding the  interaction of  squeezed vacuum  with EIT-like  media is
important not  only in filtering  applications, but also for  vapor quantum
memory protocols.

\section{Theory}

Utilizing the two-photon    formalism     developed     by    Caves     and
Schumaker~\cite{caves1985pra_two_photon_I,  caves1985pra_two_photon_II}, we
use the following expressions for amplitude ($X_+$) and phase ($X_-$) quadrature operators
\begin{eqnarray}
	X_+ = \frac{a(\Omega)+a^\dagger(-\Omega)}{\sqrt{2}}, \\
	X_- = \frac{a(\Omega)-a^\dagger(-\Omega)}{i\sqrt{2}},
	\label{eq:op_def}
\end{eqnarray}
where $a$ and $a^\dagger$ are operators of  annihilation and creation of a photon at
sideband frequency $\Omega$ with respect to the light carrier frequency.
The quantum noise  power of the corresponding quadrature is equal to
the  variance of the quadrature ($V_\pm=\left\langle X_\pm^2\right\rangle-\left\langle X_\pm \right\rangle ^2 $).
In this normalization,  coherent unsqueezed states have 
$V_\pm=1$. This corresponds to the case of shot noise,  or the standard
 quantum limit~(SQL). A coherent squeezed state then has one quadrature
variance smaller than 1 while the product $V_+ \times V_- = 1$. Such a
squeezed coherent state however is very hard to obtain in practice, and
usually
in the lab, one measures squeezed  states where $V_+ \times V_- \ge 1$.
This is true in our experiment where quadrature noise powers are 
about -2~dB for the minimum (squeezed) noise  quadrature  and 8~dB
for maximum (antisqueezed) noise
quadrature (see for example figure~\ref{fig:attenuator}).

When  a  quantum  light  state  interacts with  a  medium which has the
complex
field transmission coefficient
\begin{equation}
	{\cal{T}}(\pm \Omega) = T_\pm e^{i \Theta_\pm}
	\label{eq:complex_transmission}
\end{equation}
due  to   the  changes   to  light  transmission   and  phase,   the  noise
levels   are  altered   according   to  the   following  equation   derived
in~\cite{mikhailov2005pra_eit_rot}
%\begin{widetext}
\begin{equation}
	\begin{pmatrix}
		V_{+_{out}}\\V_{-_{out}}
	\end{pmatrix}=
	%\begin{pmatrix}
		%\cos{\phi_+} & -\sin{\phi_+} \\
		%\sin{\phi_+} & \cos{\phi_+}
	%\end{pmatrix}
	\begin{pmatrix}
		A_{+}^{2} & A_{-}^2\\
		A_{-}^2 & A_{+}^2
	\end{pmatrix}
	\begin{pmatrix}
		V_{+_{in}}\\V_{-_{in}}
	\end{pmatrix}
		+ 
	\begin{pmatrix}
			1-(A_{+}^{2}+A_{-}^{2})\\
			1-(A_{+}^{2}+A_{-}^{2})
	\end{pmatrix}
	\label{eq:v_out}
\end{equation}
%\end{widetext}
Here,  $A_{\pm}\equiv\frac{1}{2}(T_{+}\pm   T_{-})$.
The first term of the above
equation corresponds to attenuation/absorption of the propagating field,
and the
second takes into account the vacuum state which couples in and replaces
the
absorbed input field. Due to accumulated phase shifts of the positive and
negative sidebands,  the squeezed state might experience a rotation of 
the squeezing angle by  
\begin{equation}
	\phi = \frac{1}{2} (\Theta_+ + \Theta_-).
	\label{eq:rotation_angle}
\end{equation}
For  a  symmetrical  resonance 
lineshape, the Kramers-Kronig relationships dictate that
$\Theta_+ = - \Theta_-$, thus  $\phi$ is zero and no rotation occurs.

Given the  input noise for the  squeezed and antisqueezed quadratures  of a
signal, and  knowing the  light transmission  lineshape through  EIT, 
we can use these
equations to predict the output quadrature noise levels.

\section{Experimental Setup}
%Our  method for  generating squeezed  vacuum relies  upon the  Polarization
%Self-Rotation effect (PSR) and  is fully described in \cite{MatskoNWBKR02}.
%When a  strong linearly  polarized pump beam  interacts resonantly  with an
%atomic medium,  quantum fluctuations in  the orthogonal vacuum  field cause
%slight ellipticities  which lead to  polarization rotations via  PSR. While
%the  strong linear  field  is  little affected,  the  vacuum state  becomes
%quadrature squeezed. If we define the  degree of squeezing as $r_s$ and the
%squeezing angle  as $\theta_s$,  the squeezing operator  can be  written in
%terms of the photon creation and annihilation operators as
%\begin{equation}
	%S(\xi)=\exp{\left(\frac{1}{2}\xi^*\widetilde{a}^2-\frac{1}{2}\xi \widetilde{a}^{\dagger 2}\right)}
%\end{equation}
%where  $\xi=r_s  e^{i   2  \theta_s}$.  The  variances   for  the  squeezed
%and   antisqueezed  quadratures   are  then   $e^{-2r_s}$  and   $e^{2r_s}$
%respectively~\cite{bachor_guide_2004}.

\begin{figure}[h]
	\includegraphics[width=0.9\columnwidth]{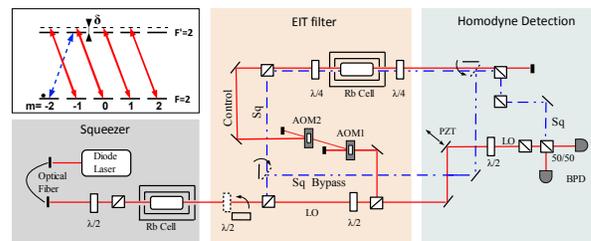}
	\caption{
		\label{fig:setup}
		Experimental setup: $\lambda/2$-~half-wave plate,
		$\lambda/4$-~quarter-wave plate, Sq-~Squeezed vacuum,
		LO-~Local oscillator, AOM-~Acousto-optical
		modulator, BPD-~Balanced photodetector. The insert shows relevant
		$^{87}$Rb sublevels and optical fields. The weak probe field is depicted
		with dashed lines, the control with solid.  $\delta$ is the two-photon detuning.
	}
\end{figure}

The arrangement of  our experiment is shown in  Figure~\ref{fig:setup}. Its three
main components are  the squeezer, where squeezed vacuum  is generated,
the EIT cell, where filtering occurs, and the balanced homodyne detector,
which boosts the optical quantum noise above the electronic dark noise
allowing it to be measured.  
Both the squeezer and EIT filter contain a Pyrex cell (length 75~mm)
containing  isotopically-enriched  $^{87}$Rb  vapor  surrounded  by  three
layers of  $\mu$-metal magnetic shielding.  The squeezing cell contains 
pure $^{87}$Rb vapor while the EIT cell contains an additional 2.5~T Neon buffer gas.

Our  method for  generating squeezed  vacuum relies  on the  polarization
self-rotation effect (PSR) suggested  in~\cite{MatskoNWBKR02}. 
This technique has been successfully demonstrated
by several groups to 
date~\cite{ries_experimental_2003, grangier2010oe, lezama2011pra,
mikhailov2008ol, mikhailov2009jmo, mikhailov_psr_mot2011, mikhailov2011jmo,
mikhailov2012sq_magnetometer, mikhailov2012sq_pulses}.  
When a  strong linearly  polarized pump beam  interacts resonantly  with an
atomic medium,  quantum fluctuations in  the orthogonal vacuum  field cause
slight ellipticities  which lead to  polarization rotations via  PSR. While
the  strong linear  field  is  little affected,  the  vacuum state  becomes
quadrature squeezed.

Our squeezer apparatus is the same as in our previous
experiments~\cite{mikhailov2012sq_pulses,mikhailov2012sq_magnetometer}
We start  with the output  of a
DL100 Toptica external cavity semiconductor  laser tuned to the D$_1$
line  $F_g=2\rightarrow  F_e=2$  transition  of  $^{87}$Rb.  A  single-mode
polarization-maintaining  fiber outputs  a  clean  symmetric Gaussian  mode
which passes through  a high quality Glan-laser polarizer  to ensure linear
polarization  of  the  beam.  The  beam then  focuses  to a
100 $\mu$m diameter beam waist inside  the  first
$^{87}$Rb vapor cell which acts as our squeezing medium.

For squeezing conditions, we choose an atomic density of
$2.9\times10^{11}$~$\frac{\text{atoms}}{\text{cm}^3}$ (corresponding to the cell
 temperature of 66$^\circ$C)
and input laser powers  ranging from 10-20~mW.  We measure  noise suppressions of
up  to  2~dB  for  the  squeezed  quadrature, as  well  as  several  dB  of
antisqueezing. We separate  this squeezed vacuum field  from the orthogonal
strong pump laser field  with  a polarizing  beamsplitter  (PBS) after the
squeezer. The  squeezed
vacuum field can  be directed straight into the detection  optics
bypassing the EIT filter cell to check
the prepared squeezed  noise levels, or it can be  directed through the EIT
vapor cell and then to detection.

To measure the noise spectrum of the squeezed/antisqueezed vacuum state, we
mix it with a  local oscillator (LO) beam on a 50/50  beam splitter as part
of a standard homodyne detection scheme  and send the resulting signal to a
spectrum analyzer~(SA). The local  oscillator is derived from the  original  squeezer pump  laser  field  with its
polarization
rotated  so  that  it
matches  that of the  squeezed vacuum. With  this experimental
setup, we are able to change the  path length traveled by the LO by
scanning the PZT voltage to achieve
different phase shifts of the LO compared with the squeezed field, and thus
measure noise  levels of the  squeezed and antisqueezed  noise quadratures.
Detection requires spatial  mode-matching   of the squeezed
beam  and local  oscillator  on the  photodiodes and  we  achieve a  $97\%$
interference fringe visibility.  The balanced  photodetection  (BPD)  is made  up
of  two  matched Hamamatsu  photodiodes  with  better than  $95\%$  quantum
efficiency.

In each measurement, we can compare the original squeezing levels with
those
modified by the EIT filter by sending the squeezed  vacuum around the EIT cell instead  of through it using flipper mirrors in the beam path. We
can also compare the measured noise  levels to the shot noise by completely
blocking the squeezed vacuum state, thereby replacing it with normal vacuum
which combines with the LO.

To facilitate EIT,  we split the power  of the strong laser  field from the
output of the squeezer  and use part of it as the  EIT control field, which
overlaps and
propagates almost colinearly with the squeezed  vacuum through the EIT vapor cell.
The quarter-wave plates on 
either  side of  the EIT  cell were  used
to convert the polarizations of the squeezed and control 
light fields  to circular and orthogonal polarizations while
traveling through the EIT cell, and then back to linear after the
cell.  We then separate the squeezed beam from the control with two
polarizing beam splitters (PBS) to improve
polarization separation,  and send it to the homodyne detector.
To further reduce the EIT control field influence, 
we introduced a slight angular misalignment between the beams so most of
the control field misses our photodiodes and thus does not introduce large
background light levels.
Since the circularly polarized control field is very strong,
it optically pumps most of the atoms into F=2 m=-2 ground sublevel and
effectively creates a single $\Lambda$ light level configuration with the
squeezed field and Zeeman sublevels of the Rb (see insert in the figure \ref{fig:setup}).

To characterize an EIT resonance we send a weak coherent probe field
 (instead of a squeezed field) into the EIT cell by introducing a
half-wave plate after the squeezer.
Since the EIT signal depends on the two-photon detuning, we sweep
this detuning. This is accomplished by two different, but essentially
equivalent methods.
In first case,
we  change  the  detuning  of  the  control  field  with  two
acousto-optical modulators  (AOMs), taking  the negative first and first
order  beams, shifted
by  -80~MHz and  $80+\delta$~MHz  respectively. The resulting control
field is detuned from the probe by $\delta$, and we have full control
over the two-photon detuning. We find that due to the AC-Stark shift induced by the
control field, the EIT
resonance is centered around 900~kHz two-photon detuning and we were not
able to fully filter the control field out, resulting in a large beat-note
resonance between the LO and the control field on our noise spectrum. We
removed
points around this resonance from our noise  spectra (see figure~\ref{fig:attenuator}b).
For subsequent measurements, we used a second method of sweeping the EIT
resonance. With the AOMs, we detuned the control field by
5~MHz and additionally introduced a magnetic field in the direction of
light propagation in the EIT cell to compensate for this detuning shift. A
calibrated sweep of this magnetic field corresponds to a
change of the two-photon detuning. With this method, the LO-control field
beat-note was placed outside of our detection band which improves the
measured noise spectra by removing the large resonant peak (see 
spectra on figures \ref{fig:filter}b and \ref{fig:rotator}b). 

During the noise EIT filter measurements, we fixed the two-photon detuning $\delta$ on top of the EIT resonance,
and the overall shape of the transmission resonance vs two-photon
detuning is shown
with respect to this fixed detuning (see figures
~\ref{fig:attenuator}a, \ref{fig:filter}a, and \ref{fig:rotator}a). In
this case, positive and negative frequency transmissions reflect the
absolute value of ${\cal{T}}(\pm \Omega)$ in
 equation~\ref{eq:complex_transmission}.
I.e. we directly measure $T_\pm$. We fit the transmission measurements to 
the following empirical function suggested in~\cite{knappe2003}:
\begin{equation}
T_{\pm}=A\frac{\Gamma ^{2}}{\Gamma ^{2} + (\delta_0 \pm
\Omega)^{2}}+B\frac{\Gamma(\delta_0 \pm \Omega)}{\Gamma ^{2} + (\delta_0 \pm \Omega)^{2}} +C
\end{equation}
Here the first and second terms are the symmetric and anti-symmetric
Lorentzian and the last constant term represents residual absorption of the
light due to incoherent processes; $\Gamma$ is the effective half width
half maximum  of the resonance; $\delta_0$ is the shift of the EIT
resonance with respect to squeezed vacuum field 
(Essentially zero, as we keep the two-photon detuning on top of the EIT resonance);
$A$, $B$, and $C$ are the fitting parameters. Once we have the numerical
expression for $T_\pm$, we
input these transmission coefficients into equation~\ref{eq:v_out} to predict
the output noise level. 
We did not measure the sideband phase
lag $\Theta_\pm$ in
our experiments, and so neglected the squeezing angle
rotation in our calculations.

\section{Experimental observations}

\begin{figure}[h]
	\includegraphics[width=0.9\columnwidth]{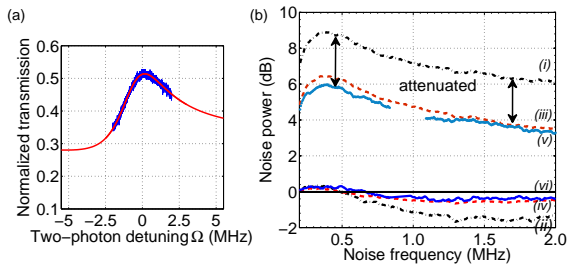}
	\caption{
		\label{fig:attenuator} 
		(a) EIT lineshape: Solid line shows fit.  Peak transmission=~$52\%$, FWHM=~4~MHz,  control
		 power=~4.2~mW, EIT cell temperature $T_{EIT}$=~$46^{\circ}$~C.  (b)
		 Quadratures noise power spectra. 		{(i) input  max.
		noise, (ii)  input min.  noise, (iii) expected  max. noise,  (iv)
		expected
		min. noise, (v) measured max. noise, (vi) measured min. noise.}
		Squeezer pump
		power=~21.6~mW, squeezing cell temperature $T_{sq}$=~$59^{\circ}$~C.
		We  removed    data points in the output noise between  0.8 and  1.1~MHz due  
		to a  large spike  caused by  the beatnote between the local oscillator
		and the control field which was detuned by 900~kHz and leaked into the 
		detection. 
		}
\end{figure}

\begin{figure}[h]
	\includegraphics[width=0.9\columnwidth]{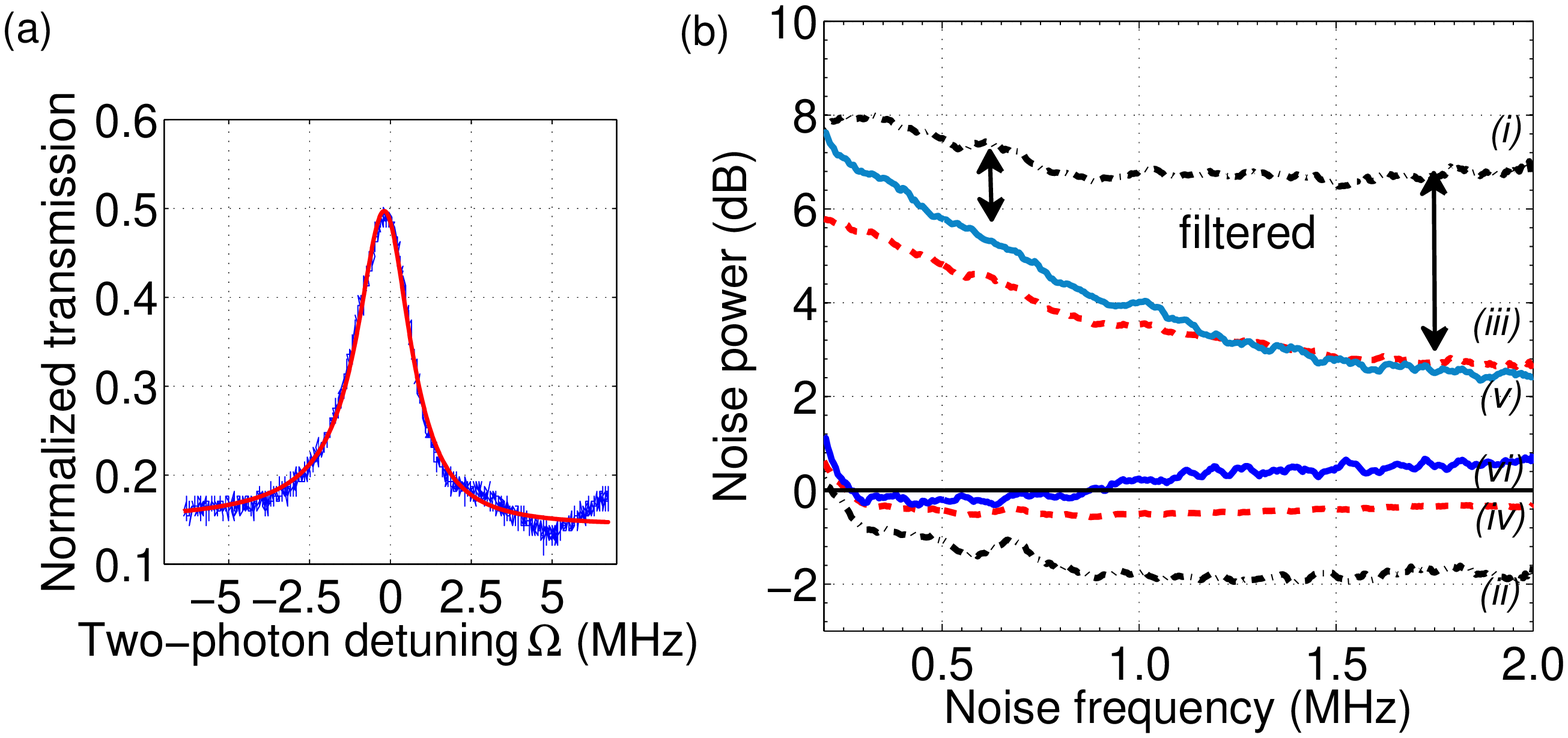}
	\caption{
		\label{fig:filter}
		(a) EIT lineshape: Solid line shows fit.  Peak transmission=~$50\%$, FWHM=~2~MHz,  control
		power=~3.8~mW, EIT cell temperature $T_{EIT}$=~$50^{\circ}$~C.
		(b)  Quadratures noise power spectra. 
  	{(i) input  max.
		noise, (ii)  input min.  noise, (iii) expected  max. noise,  (iv)
		expected
		min. noise, (v) measured max. noise, (vi) measured min. noise.}
		Squeezer pump power=13~mW, squeezing cell
		temperature $T_{sq}$=~$57^{\circ}$~C.
	}
\end{figure}

The influence  of the  atomic medium  on the squeezed  vacuum state  can be
observed by  comparing the input  minimum and  maximum noise levels  to the
levels measured after interaction. By attenuating the control field and
thus decreasing the power broadening of the EIT resonance,
or by slightly changing the control field alignment, we can
narrow the  EIT linewidth and  change the  transmission window used  in the
experiment. As  an example  of squeeze amplitude  attenuation, we  show two
noise spectra    and their  associated EIT  transmission curves  in figures
\ref{fig:attenuator} and \ref{fig:filter}.  In figure \ref{fig:attenuator}(b), we
start
off  with a  squeezed vacuum  showing up  to 1.5~dB noise  suppression and
nearly 9~dB of excess, antisqueezed  noise. Any frequency dependence of the
input noise  levels, we attribute to  laser noise which was  not completely
subtracted by the balanced photodiodes. This prevents us from detecting the
best squeezing  at the lowest  noise frequencies,  but this problem  can be
alleviated  by better mode matching  and alignment of the LO and the squeezed beam at the BPD,
with improved beam pointing stabilization. We note
that with this exact same squeezer, but using a modified squeezing
detection scheme
(not suitable for this experiment), we were able to generate squeezing to
frequencies as low as 200~Hz~\cite{mikhailov2012sq_magnetometer}. 
For the first measurement, we make the  transmission curve 
rather broad, with full width half maximum (FWHM) of the resonance $>4$~MHz
(see figure \ref{fig:attenuator}(a)),
and with a fairly small contrast between the  peak transmission of $52\%$
and the
background transmission of 28\%. As a result,  in figure
\ref{fig:attenuator}(b), the
output
noise levels  are uniformly attenuated due to  the light absorption, but  there is no
visible frequency-dependent filtering  of the noise, since in the detection
bandwidth of 2~MHz, transmission for all sidebands is almost the same.  
We also calculate the expected filtered noise spectra based on
equation~\ref{eq:v_out} and 
transmission coefficients ($T_\pm$) extracted from the fit of the EIT
transmission data (figure \ref{fig:attenuator}(a)).  We see a very good
match between the theoretical prediction and the experimental data.
The output  noise follows along  the same
shape  as the  input noise  close to  the predicted  noise levels,  without
changes  in its  frequency-dependence.  We  removed    data
points in the output noise between  0.8 and  1.1~MHz due  
to a  large spike  caused by  the beatnote between the local oscillator
and the control field which was detuned by 900~kHz and leaked into the 
detection. 

Note  the difference  in figure  \ref{fig:filter}.  Here, as  shown in  
figure~\ref{fig:filter}(a), we narrow the EIT transmission window
to about 2~MHz FWHM and increase the contrast between maximum
and background transmission.  Now, with similar input  squeezed (-2~dB) and
antisqueezed (8~dB) noise  levels, the output noise  shows marked frequency
dependence. At  lower frequencies where  transmission is at a  maximum, the
output noise levels are closer to the inputs, but at higher frequencies, we
see more and more attenuation due to the light absorption at the wings of
the EIT resonance.  
This data set shows the simple use of the EIT window as a low-pass
filter.  The effects of the filter are most easily observed in the
antisqueezed noise quadrature due to the high amplitude, starting with
8~dB of excess noise.  The squeezed quadrature also appears to follow
the shape of the filter, but due to extra noise raising the noise floor,
this minimum level rises above shot noise rather than settling closer to
it.  

We attribute this extra noise to the several potential  sources.
First, our numerical prediction model assumes that there is no 
squeezing angle rotation influencing the noise in
figures \ref{fig:attenuator} and \ref{fig:filter}.
This is clearly an oversimplification, because a small visible asymmetry
of the EIT resonance in figure  \ref{fig:filter}
dictates, according to the Kramers-Kronig relations, that some frequency
dependent
rotation should be present, which may show up as a deviation from the
predicted noise
levels. 
A second possibility is the simplicity of our model, which treats
the EIT resonance as a passive absorptive filter and disregards
the back action of light noise onto the atoms as well as the atomic noise
contribution. 
This simple approach may be successful up to a point, but
could lead to deviations from experiment when excess noise contributions 
become sizable.  We note that in this experiment, the noise level
resulting from blocking the squeezed probe before the EIT cell was 
identical to that seen when the probe was blocked after the atoms and 
just before detection (shot noise).  This leads us to believe the 
atomic noise contribution is small in this case, and that most of the
excess noise must then be due to back action of the light noise.
Lastly, as mentioned, any laser noise which is imperfectly balanced by the 
homodyne detector can raise the noise floor and add apparent frequency dependence.

\begin{figure}[h]
\includegraphics[width=0.9\columnwidth]{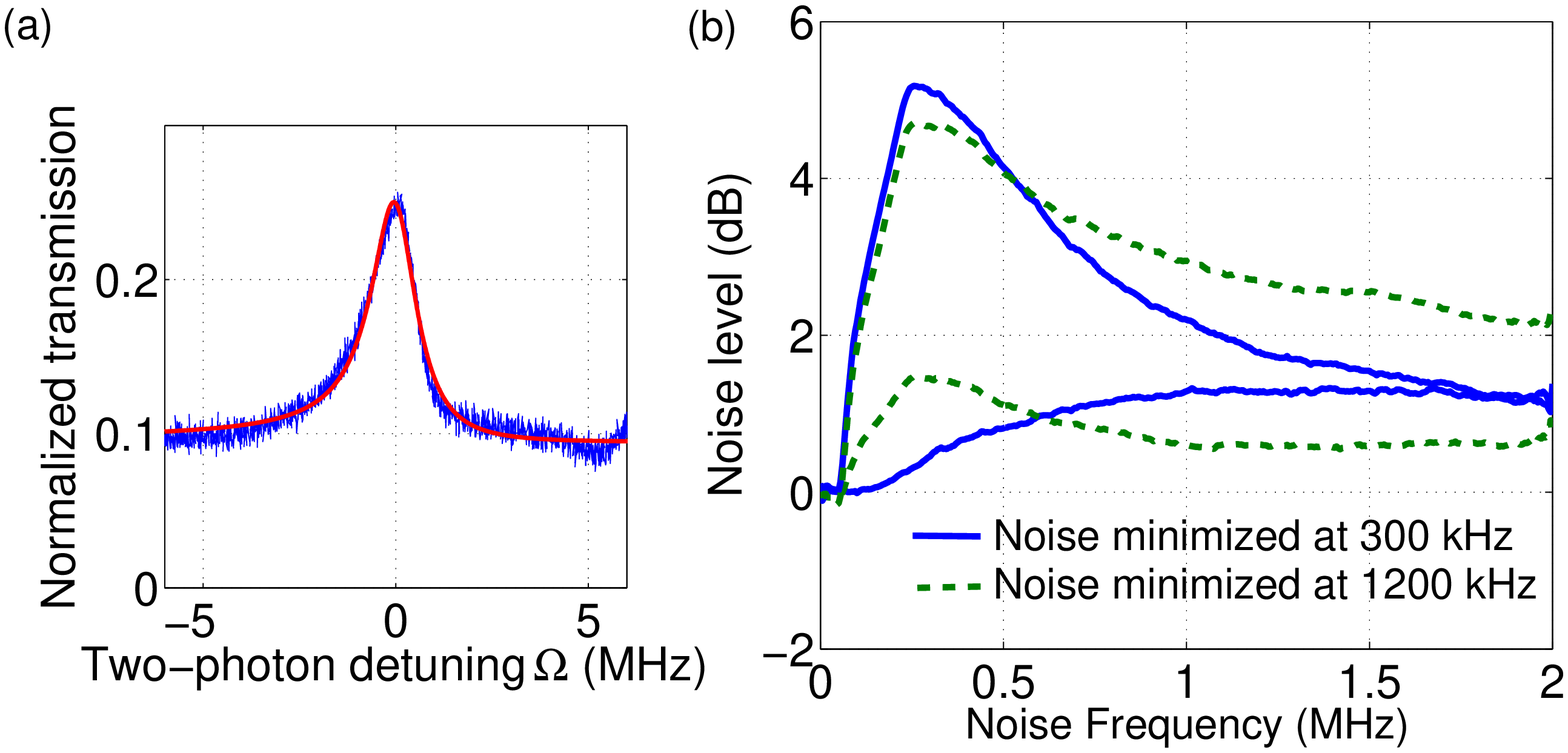}
\caption{
\label{fig:rotator}
(a) EIT lineshape: Solid line shows fit. Peak transmission=~$25\%$, FWHM=~1.4~MHz,  control
		power=~2.3~mW, EIT cell temperature $T_{EIT}$=~$50^{\circ}$~C.
(b) Quadratures noise power spectra: with noise power minimization at 300~kHz
(solid-blue line) and at 1.2~MHz (dashed-green line).
Squeezer pump power=15~mW, squeezing cell temperature $T_{sq}$=~$57^{\circ}$~C.
}
\end{figure}

Figure~\ref{fig:rotator} depicts the very interesting effect
of frequency-dependent squeeze angle rotation.
Here, because  of the asymmetry of  the EIT lineshape,
there is a resulting phase shift between the left and right noise
sidebands leading
to a rotation of
the squeezing angle, which now changes with frequency. We see that the
LO  phase 
chosen  for the  best noise suppression at one  noise frequency  is not 
the phase
giving the minimum noise level at  all frequencies. This indicates that the
squeezing angle has actually become  frequency-dependent as it rotates with
frequency, requiring different phases for different noise 
frequencies in order to measure the maximum squeezing.  Note that the 
noise spectrum resulting from choosing the proper phase  at a 
lower frequency (300~kHz), looks very different from the result
when the minimum noise is found by choosing the phase angle at a 
higher noise frequency (1200~kHz). 

To the best of our knowledge this is first reported measurement of the
squeezing angle rotation done with atoms which was previously theoretically
predicted in~\cite{mikhailov2005pra_eit_rot}. Until now the only
successfully  reported way to rotate the squeezing angle was with
cavities~\cite{schnabel2005pra_squeezing_control}. Unfortunately, our data
has a lot of excess noise and sub shot noise reduction did not survive
after the passage through EIT.

We note that for experimental conditions corresponding to figures
\ref{fig:attenuator} and \ref{fig:filter}, we did not observe such
rotation. Here the squeezing angle seems to show good frequency independence.

\begin{figure}[h]
\includegraphics[width=0.5\columnwidth]{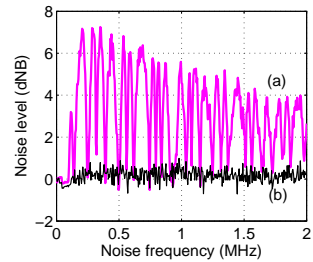}
\caption{
\label{fig:control_off2}
Noise power spectrum with control field set to 6.9~mW (a) and blocked (b).
LO phase angle is continuously scanned.
}
\end{figure}

We also demonstrate the capability to completely replace the squeezed state
with unsqueezed ordinary coherent vacuum state by changing the EIT 
media to a strong absorber by switching off the control field.
The final plot,  figure \ref{fig:control_off2}, shows the output noise  levels
of the quantum state after the EIT filtering cell, first~(a),
while the  control  field is on and we
sweep the local oscillator phase while recording the noise spectrum, 
and second~(b), in the same situation but with the control beam
completely blocked.  Note  that while  the control beam is on, we see
high phase-dependent noise levels and the squeezed vacuum is transmitted through
the atoms.  However, with the control off, we do 
not have EIT conditions  and our vacuum  state is absorbed by  the atoms.
Thus, the output noise level corresponds to shot noise, identical to the
noise spectrum generated by blocking the squeezed path just before the BPD.
As expected,  the quantum noise is  not transmitted through the
atomic medium, so the noise  level returns to shot noise and does not
depend on the LO phase.
Such a switchable filter can be of interest for quantum repeaters and
quantum memory protocols.

\section{Conclusion}

We  have  shown   an  experimental demonstration  of  EIT used for frequency-dependent
squeeze amplitude   attenuation  
of a quantum  squeezed vacuum field using only atomic vapors.  In our
experiment, the atoms act
as a low-pass filter for squeezed and antisqueezed noise.  
The relative ease of controlling the transmission window
in vapor cell experiments makes this method very promising for
creating several different types of noise filters for squeezed
vacuum.  This controllable squeezed vacuum source may be 
easily incorporated into precision measurement experiments
due to its simple, all-atomic design, and low loss.

We also observe an apparent frequency-dependent rotation of the squeezing angle
as the vacuum propagates through EIT.  This effect is likely
due to the asymmetry of the transmission window and could
also be used to create more complicated noise filters as well as
to match the squeezing angle to the ponderomotive squeeze angle
caused by radiation pressure in high-powered interferometers~\cite{mikhailov2005pra_eit_rot}.

We note the less well-understood excess noise sources which couple into this 
experiment and degrade the noise suppression. Their contribution cannot be
explained by simple treatment of the EIT resonance as a media with complex
transmission coefficients. A full quantum mechanical
treatment of the light-atom interaction would be needed to fully describe
the extra noise sources.
These dynamics of quantum noise,
along with any excess noise sources, will be important to any precision 
measurement or quantum memory experiment which uses the interaction
of squeezed light with an atomic medium.

%\section{Acknowledgments}

%\bibliographystyle{apsrev4-1} % no need to do it by hands with natbib > 8.31b
\bibliographystyle{tMOP}

\end{document}